\documentclass[twocolumn]{aa}
\usepackage{graphicx}
\usepackage{txfonts}
\begin{document}
   \title{A Model for Quasi-Spherical Magnetized Accretion Flow}

   \author{Mohsen Shadmehri}

   \offprints{Mohsen Shadmehri}

   \institute{Department of Physics, School of Science, Ferdowsi University,
Mashhad, Iran\\
              \email{mshadmehri@science1.um.ac.ir}}

   \date{Received ...; accepted ...}

   \abstract{
   A model for axisymmetric magnetized accretion flow is proposed. The dominant mechanism of energy dissipation is
assumed to be the magnetic diffusivity  due to turbulence in the
accretion flow. In analogy to the advection-dominated accretion
flow (ADAF) solutions, a constant fraction of the resistively
dissipated energy is stored in the accreting gas and the rest is
radiated. The general self-similar solutions which describe a
resistive and {\it nonrotating} flow with purely poloidal magnetic
field is first introduced. The radial dependence of physical
quantities  is identical to viscous ADAF solutions. Although the
main focus of this study is nonrotating magnetized accretion
flow, for rotating flow with both poloidal and toroidal components
of magnetic field we find the radial scaling of solutions similar
to the nonrotating case. We show that the accretion and the
rotation velocities are both below the Keplerian rate,
irrespective of the amount of cooling. We show that the set of
the equations is reduced to a second order differential equation
for a nonrotating flow. The geometrical shape of the disk changes
depending on the fraction of the resistively dissipated energy
which is stored in the accreting gas. However, there is a hot
low-density gas above the disk in almost all cases. The net
accretion rate is calculated for a set of illustrative parameters.
   }

   \maketitle
%

\section{Introduction}

   The importance of the accretion processes has been recognized in
connection with the physics of many astrophysical systems such as
protosteller disks (e.g., Mineshige \& Umemura 1997; Tsuribe 1999;
Calvet, Hartmann \& Strom 2000), disks around black holes or
compact objects  or disks at center of the active galactic nuclei
(e.g., Lynden-Bell \& Rees 1971; Kato, Fukue \& Mineshige 1998).
According to their geometrical shapes, accretion disks are
generally divided into thin and thick disks. A mechanism for
transport of angular momentum is another key ingredient in theory
of accretion processes and many theoretical uncertainties remain
about its nature. However, the theory of thin disks is well
understood based on a pioneer work by Shakura \& Sunyaev (1973).
In this theory it is assumed that there is efficient radiation
cooling in the flow.

During recent years another type of accretion disks has been
studied, in which the energy released through viscous processes
in the disk may be trapped within the accreting gas (see, e.g.,
Ichimaru 1977; Narayan \& Yi 1995; hereafter NY). In this kind of
flow which is known as adection-dominated accretion flow (ADAF),
the accreting gas has a very low density and is unable to cool
efficiently. A full analysis of the dynamics of the solution was
presented in an important paper by NY. Since both the angular
momentum and the energy dissipation in the flow is undertaken by
the turbulent viscosity similar to the standard $\alpha$ model, we
can call such flows "viscous" ADAF models.

There are growing evidences that the accretion flows in many
astrophysical systems involve magnetized plasma. Accretion disks
containing magnetic fields have been studied by many authors
(e.g., Ogilvie 1997; Hawley 2001; Casse \& Keppens 2002). However,
the standard viscous ADAF model completely ignores the presence
of an ordered global magnetic field and it is assumed the
turbulent magnetic field contribute a constant fraction of the
total pressure (NY). However, we can expect accretion flows, in
which angular momentum is transported by a global magnetic field
and the energy is released by Joule heating. This possibility has
been addressed by Kaburaki (2000). He presented a set of
analytical solutions for a fully advective accretion flow in a
global magnetic field and the conductivity is assumed to be
constant for simplicity.

As for the accretion flows in presence of magnetic field,
Schwartzman (1971) was the first to point out the importance of
the magnetic field in an accretion process. He proposed a
hypothesis of equipartition between the magnetic and kinetic
energy densities and this picture as mentioned above is usually
accepted in modern picture of viscous ADAF models (e.g., NY).
More accurate study was done by Bisnovatyi-Kogan \& Ruzmaikin
(1974), where exact nonstationary solution for a field
amplification in the radial accretion flow was also obtained.
Also, Bisnovatyi-Kogan and Lovelace (2000) suggested that recent
papers discussing ADAF as a possible solution for astrophysical
accretion should be treated with caution, particularly as for
ignorance of the magnetic field. While they obtained a solution
for time-averaged magnetic field in a quasispherical accretion
flow, an analysis of energy dissipation and equipartition between
magnetic and flow energies has been presented (Bisnovatyi-Kogan
and Lovelace 2000). In our study, we follow similar ideas in
order to study magnetized accretion flow in which magnetic energy
dissipation is stored in the flow.

Dynamics of radiatively inefficient accretion flows has been the
subject of many studies during recent years (e.g., Igumenshchev \&
Abramowicz 1999, 2000; Stone, Pringle \& Begelman 1999; McKinney
\& Gammie 2002). In these viscous accretion models, the behaviour
of the flow depends on the standard dimensionless viscosity
parameter $\alpha$. On the other hand, as rotation is allowed,
such flows become subject to strong convection with completely
different flow pattern and radiative properties (Igumenshchev \&
Abramowicz 1999; Stone, Pringle \& Begelman 1999). Numerical MHD
simulations of radiatively inefficient flows have been done
recently by many authors (e.g., Hawley 2001; Machida, Matsumoto \&
Mineshige 2001; Casse \& Keppens 2002). But most of these studied
have neglected the resistive terms in the MHD equations or  the
resistivity has been considered only in the induction equation
without accounting the corresponding dissipation in the energy
equation.

 In this paper, we want to explore how the structure of a
steady-state thick disk depends on its resistivity by solving the
two-dimensional MHD equations for disks which have the
simplifying feature that they are self-similar in radius. We will
consider the general problem of accretion flow where  the field
annihilation is approximated by a finite "turbulent" conductivity
and the viscosity of the fluid is completely neglected in order
to construct a fully resistive model for accretion within a
global magnetic field. We restrict our attention to the
nonrotating accretion flow, in which the flow contains a purely
poloidal magnetic field, and the general case will be reported in
future. However, we will find the radial scaling of self-similar
solutions for the general case where the fluid is rotating and it
contains both poloidal and toroidal components of the magnetic
field.

This paper is organized as follows. In section 2 the general
problem of constructing a model for resistively accretion flow is
defined. The self-similar solutions are presented in section 3,
and the effects of the input parameters are examined. The
implications of this analysis are summarized in section 4.


\section{Formulation of the Problem}

As stated in the Introduction, we are interested in constructing
a model for describing magnetized accretion disks. The macroscopic
behavior of such flows can be studied by MHD approximation. For
simplicity, the self-gravity and general relativistic effects
have been neglected. The first assumption is true as long as the
total mass of the disk is small compared with the product of the
mass of central object and the angular thickness of the disk.
Thus, the basic equations of our problem are written as follows:
\begin{equation}
\frac{\partial\rho}{\partial t}+\nabla.(\rho{\bf v})=0,
\label{eq:cont}
\end{equation}
\begin{equation}
\rho(\frac{\partial{\bf v}}{\partial t}+({\bf v}.\nabla){\bf
v})=-\nabla p -\rho \nabla\Psi + \frac{1}{4\pi}{\bf
J}\times\bf{B}, \label{eq:motion}
\end{equation}
\begin{equation}
\frac{\partial{\bf B}}{\partial t}=\nabla\times({\bf v}\times{\bf
B }-\eta {\bf J}), \label{eq:induction}
\end{equation}
\begin{equation}
\frac{\rho}{\gamma-1}\frac{d}{dt}(\frac{p}{\rho})+p \nabla.{\bf
v}=Q_{\rm diss}-Q_{\rm cool} \label{eq:energy}
\end{equation}
\begin{equation}
\nabla . {\bf B}=0, \label{eq:divb}
\end{equation}
where $\rho$, ${\bf v}$, $p$, $\Psi$ are the density, the
velocity, the pressure and the gravitational potential due to a
central object, respectively. We assume that the mass of central
object is $M$ and so, the gravitational potential becomes
$\Psi=-GM/r$ where $r$ is the spherical radial coordinate. Also,
${\bf J}=\nabla\times {\bf B}$ is the current density and $\eta$
represents the magnetic diffusivity. The term on the right hand
side of the energy equation, $Q_{\rm diss}$, is the rate of
heating of the gas by the dissipation and $Q_{\rm cool }$
represents the energy loss through radiative cooling and $\gamma$
is the adiapatic index. We assume that the difference between
energy dissipation and radiative cooling is a fraction $f$ of
energy dissipation, i.e. $Q_{\rm diss}-Q_{\rm cool}=f Q_{\rm
diss}$ (e.g., NY). The parameter $f$ measures the degree the flow
is advection-dominated and in general it may depends on the
details of the heating and cooling mechanisms and will vary with
both $r$ and $\theta$. However, we assume a constant $f$ for
simplicity. Clearly, the case $f=1$ corresponds to the extreme
limit of no radiative cooling and in the limit of efficient
cooling, we have $f=0$.

We assume that the disk is stationary and axisymmetric. There are
numerous studies, in which magnetized equilibrium structures have
been investigated via analytical or numerical methods (e.g.,
Igumenshchev \& Abramowicz 1999, 2000; Stone, Pringle \& Begelman
1999; Shalybkov \& Rüdiger 2000; Ogilvie \& Livio 2001; McKinney
\& Gammie 2002;  Rüdiger \& Shalybkov 2002). This paper will be
restricted to nonrotating accretion flow under purely poloidal
magnetic fields configuration.  The whole magnetic  field is
maintained by the electric current induced in the accretion
region, and the solutions are determined only by the assumption
of self-similarity. According to the anti-dynamo theorem (e.g.,
Cowling 1981), such a configuration cannot  be maintained in a
stationary state by an axisymmetric flow in the presence of the
resistive dissipation. We can resolve this criticisms in two
forms: For many problems, the magnetic dissipation time is very
long, much longer than the age of the system.  For such problems,
one assumes the field to have been generated in the distant past,
and restricts one's attention to its effect on current dynamics,
equilibrium, stability, etc.

For other problems, the dissipation time is not longer than the
age of the system.  In such cases (as in this study), one is
making some implicit assumption about the regeneration of the
field by (dynamo) processes outside of the computational regime
of the problem at hand (e.g., on a 3-D time-dependent microscale,
or beyond the boundaries of the formal computation).  Whether
these assumptions are consistent with the computational details
is, of course, a function of how good is the author's physical
intuition, since such assumptions lie beyond the physical and
mathematical equations that are actually posed and solved.

There many studies, in which the time-averaged physical
quantities have been analyzed so that one can still talk about
stationary state and this approach has been widely used in
literature (e.g., Lovelace, Wang \& Sulkanen 1987; Lovelace,
Romanova \& Newman 1994;  Shalybkov \& Rüdiger 2000; Ogilvie \&
Livio 2001; Rüdiger \& Shalybkov 2002). In this study the
physical quantities such as velocity $\bf v$, magnetic field $\bf
B$ etc. denote the time-averaged values and in general, the
instantaneous value of each physical variable consists of a
stationary part and a fluctuating component. The influence of the
fluctuating parts is included crudely, by modifying the transport
coefficients which would apply in the absence of fluctuations.
Thus, we neglect the time average of fluctuating parts of
velocity multiply magnetic field and the generation of magnetic
field in the disk by dynamo activity is not included (e.g.,
Pudritz 1981). Since we are interested in nonrotating
quasispherical accretion, one may ask about the instantaneous
magnetic field lines. Recently, Bisnovatyi-Kogan and Lovelace
(2000) proposed a model for magnetized quasispherical accretion,
in which a sketch of the instantaneous poloidal magnetic field
lines and the time-averaged flow velocity has been shown in
Figure 1 of their work. As for the accretion flow within global
poloidal magnetic field, we are assuming the same configuration.

We work in spherical polar coordinate $(r,\theta,\varphi)$ with
the origin on the central object. We assume that the net mass
accretion rate $\displaystyle\dot{M}=-\int 2\pi r^2 \sin\theta
\rho v_{\rm r} d\theta$ is independent of radius $r$. Since $r^2
\rho v_{\rm r}$ is independent of $r$ and the disk is stationary
and axisymmetric, the continuity equation (\ref{eq:cont}) shows
$v_{\rm \theta}=0$ and takes the form
\begin{equation}
\frac{1}{r^{2}}\frac{\partial}{\partial r}(\rho v_{\rm r}
r^{2})=0.\label{eq:continuty}
\end{equation}
The condition of constant $\dot{M}$ is similar to what has been
used by NY, however, is relaxed by some authors and they find
solutions with outflows (e.g., Henriksen \& Valls-Gaboud 1994). In
this paper, we consider the simple case of constant net mass
accretion rate at all radii. We can write the components of
equation of motion (equation (\ref{eq:motion})) in spherical
coordinates:
\begin{equation}
\rho(v_{\rm r}\frac{\partial v_{\rm r}}{\partial r}-\frac{v_{\rm
\varphi }^{2}}{r})=-\frac{\partial p}{\partial r}-\rho
\frac{GM}{r^{2}}+\frac{1}{4\pi}(J_{\rm \theta}
B_{\rm\varphi}-J_{\rm\varphi}B_{\rm\theta}),\label{eq:motionr}
\end{equation}
\begin{equation}
-\rho v_{\rm\varphi}^{2}\cot\theta=-\frac{\partial
p}{\partial\theta}+\frac{r}{4\pi}(J_{\rm\varphi}B_{\rm r}-J_{\rm r
}B_{\rm\varphi}),\label{eq:motiontet}
\end{equation}
\begin{equation}
\rho v_{\rm r}\frac{\partial v_{\rm\varphi}}{\partial r}+\rho
\frac{v_{\rm r}v_{\rm\varphi}}{r}=\frac{1}{4\pi}(J_{\rm
r}B_{\rm\theta}-J_{\rm\theta}B_{\rm r}).\label{eq:motionphi}
\end{equation}
Also, the three components of induction equation
(\ref{eq:induction}) are:
\begin{equation}
\frac{\partial}{\partial\theta}[r \sin\theta(v_{\rm
r}B_{\rm\theta} -\eta J_{\rm \varphi} )]=0, \label{eq:II}
\end{equation}
\begin{equation}
\frac{\partial}{\partial r}[r\sin\theta( v_{\rm r}B_{\rm\theta}
-\eta J_{\rm \varphi} )]=0,\label{eq:III}
\end{equation}
\begin{displaymath}\frac{\partial}{\partial r}(rv_{\rm \varphi}B_{\rm r}-rv_{\rm
r}B_{\rm\varphi})+\frac{\partial}{\partial\theta}(v_{\rm\varphi}B_{\rm\theta})-\frac{\partial}{\partial
r}(\eta r J_{\rm\theta})
\end{displaymath}
\begin{equation}
\\+\frac{\partial }{\partial\theta}(\eta
J_{\rm r})=0.\label{eq:IIII}
\end{equation}
In equations (\ref{eq:II}), (\ref{eq:III}) and (\ref{eq:IIII}),
the terms without $\eta$ represent the advection of the magnetic
field by the flow motion, while the terms containing $\eta$,
account for the diffusion of the field.

From equations (\ref{eq:II}) and (\ref{eq:III}) one can simply
deduce that $r\sin\theta( v_{\rm r}B_{\rm\theta} -\eta J_{\rm
\varphi} )=0$, or
\begin{equation}
v_{\rm r}B_{\rm\theta} -\eta J_{\rm \varphi}=0.\label{eq:IV}
\end{equation}
This important equation can be understood easily in this way: In
a stationary, axisymmetric configuration, the electric field can
be written as a gradient of a scalar potential, and its $\phi$
component (i.e., $E_{\rm\varphi}$) which is in proportion to
$v_{\rm r}B_{\rm\theta} -\eta J_{\rm \varphi}$ in our notation,
should vanish identically. This expression  appears  in the
parentheses in equations (\ref{eq:II})  and (\ref{eq:III}). So,
these equations reduce into one equation, i.e. equation
(\ref{eq:IV}), as we showed above. We will find self-similar
solutions so that satisfy into equation (\ref{eq:IV}).

The energy equation (\ref{eq:energy}) and the magnetic flux
conservation give
\begin{equation}
\rho(\frac{v_{\rm r}}{\gamma -1}\frac{\partial}{\partial
r}(\frac{p}{\rho})-\frac{p}{\rho^2}v_{\rm r}\frac{\partial
\rho}{\partial r})=f Q_{\rm diss},\label{eq:energyr}
\end{equation}
\begin{equation}
\frac{1}{r^{2}}\frac{\partial}{\partial r}(r^{2} B_{\rm
r})+\frac{1}{r
\sin\theta}\frac{\partial}{\partial\theta}(B_{\rm\theta}
\sin\theta)=0,\label{eq:divb1}
\end{equation}
where in the above equations the components of the current
density $\bf J$ are
\begin{displaymath}
\ J_{\rm
r}=\frac{1}{r\sin\theta}\frac{\partial}{\partial\theta}(B_{\rm\varphi}\sin\theta),
\hspace{0.5cm} J_{\rm\theta}=-\frac{1}{r}\frac{\partial}{\partial
r}(r B_{\rm\varphi}),
\end{displaymath}
\begin{displaymath}
\ J_{\rm\varphi}=\frac{1}{r}[\frac{\partial}{\partial
r}(rB_{\rm\theta})-\frac{\partial B_{\rm r}}{\partial\theta}].
\end{displaymath}
To close the system of equations, we should specify the
functional forms of the magnetic diffusivity $\eta$ and the
energy dissipation $Q_{\rm diss}$. In this study, we have
completely neglected the effects of viscous processes and the
attention has been restricted only to the resistive dissipation.
In other words, the accretion flow of our model  is {\it not}
infinitely conducting. For simplicity, we could assume the
resistivity to be constant (see, e.g., Kaburaki 2000). But, we
assume  the magnetic diffusivity is due to turbulence in the
accretion flow and it is reasonable to express this parameter in
analogy to the $\alpha$-prescription of Shakura and Sunyaev
(1973) for the turbulent viscosity,
\begin{equation}
\eta=\eta_{\rm 0} \frac{c_{\rm s}^2}{\Omega_{\rm K}}=\eta_{\rm
0}\frac{p}{\rho\Omega_{\rm K}},\label{eq:diffu}
\end{equation}
where $\Omega_{\rm K}=\sqrt{GM/r^3}$ is Keplerian  angular
velocity, and $c_{\rm s}$ is the isothermal sound speed. NY
applied similar functional form for kinematic coefficient of
viscosity, i.e. $\nu=\alpha(p/\rho\Omega_{\rm K})$. Exactly in
analogy to NY's prescription, we are using the above equation as
for magnetic diffusivity $\eta$. Note that magnetic diffusivity
$\eta$ is {\it not} constant and depends on the physical variables
of the flow and $\eta$ in our self-similar solutions, as we will
show, scales with radius as $r^{1/2}$. The magnetic diffusivity
has the same units as kinematic viscosity and we assume that the
magnitude of $\eta$ is comparable to that of turbulent viscosity
(Bisnovatyi-Kogan and Ruzmaikin, 1976; Parker , 1979 ), although
we have neglected kinematic viscosity. This form of scaling for
diffusivity has been widely used by many authors (e.g., Lovelace,
Wang \& Sulkanen 1987; Lovelace, Romanova \& Newman 1994; Ogilvie
\& Livio 2001; Rüdiger \& Shalybkov 2002). Due to arbitrary
$\eta_{\rm 0}$ it is not so important which velocity is used for
the scaling. For example, in some studies Alfv\'{e}n velocity is
used for the velocity in the above scaling (e.g., Fendt \&
Cemeljic 2002 ), while in other works still sound speed is
applied even in the magnetized case (e.g.,
 Lovelace, Romanova \& Newman 1994; Rüdiger \&
Shalybkov 2002). Nevertheless, the sound speed, it seems, has an
advantage; because it is the largest intrinsic velocity parameter
in the turbulent plasma, and may be used universally, with or
without magnetic field. In this study, following many previous
studies, we use the above parameterized form for the diffusivity.

For the heating term, $Q_{\rm diss}$, we may have two sources of
dissipation. Magnetic reconnection may lead to energy release.
Also, we can consider the viscous and resistive dissipations due
to a turbulence cascade. In this study, only the resistive
dissipation has been considered as the dominant process:
$\displaystyle Q_{\rm diss }=\frac{\eta}{4\pi}J^2$.

Now, we have constructed our model and the main equations of the
model are equations (\ref{eq:continuty}), (\ref{eq:motionr}),
(\ref{eq:motiontet}), (\ref{eq:motionphi}), (\ref{eq:IIII}),
(\ref{eq:IV}), (\ref{eq:energyr}) and (\ref{eq:divb1}). In the
next section, we will present self-similar solutions of these
equations.

\section {Analysis}
\subsection{Self-Similar Solutions}
To better understand the physics of our  accretion flow in a
global magnetic field, we seek self-similar solutions of the
above equations. Of course, this method is familiar from its wide
applications to the full set of equations of MHD in many research
fields of astrophysics. As long as we are not interested in
boundaries of the problem, such solutions can accurately describe
the behavior of the solutions in an intermediate region far from
the radial boundaries.

Before presenting the self-similar solutions, we introduce a
convenient functional form for the magnetic field. Generally, we
may have both poloidal and toroidal components of the magnetic
field. We can define a positive-definite magnetic flux function
$\Phi(r,\theta)$ from which we may obtain the poloidal components
of the magnetic field:
\begin{equation}
\ {\bf B}=\frac{1}{2\pi}\nabla\times(\frac{\Phi}{r\sin\theta}{\bf
e}_{\varphi})+ B_{\varphi}{\bf e}_{\varphi}.
\end{equation}
We can easily show that by integrating the above equation over
the circular area $\pi r^2$ at plane of $\theta=\pi/2$, the
result, i.e. $\Phi(r, \pi/2)$, is the defined magnetic flux. This
equation  automatically satisfies the equations (\ref{eq:divb1}).

Writing the equations in a non-dimensional form, that is, scaling
all the physical variables by their typical values, brings out
the non-dimensional variables. We can simply show that a solution
of the following form satisfies the equations of our model:
\begin{equation}
\rho(r,\theta)=\rho_{\rm 0}\frac{R(\theta)}{(r/r_{\rm 0})^{3/2}},
\end{equation}
\begin{equation}
\ p(r,\theta)=p_{\rm 0}\frac{P(\theta)}{(r/r_{\rm 0})^{5/2}},
\end{equation}
\begin{equation}
\ v_{\rm r}(r,\theta)= r \Omega_{\rm K}(r) V(\theta),
\end{equation}
\begin{equation}
\ v_{\rm \varphi}(r,\theta)=r\Omega_{\rm K}(r)\Omega(\theta),
\end{equation}
\begin{equation}
\ B_{\rm r}(r,\theta)=\frac{B_{\rm
0}}{2\pi\sin\theta}\frac{d\phi(\theta)}{d\theta}\frac{1}{(r/r_{\rm
0})^{5/4}},
\end{equation}
\begin{equation}
\ B_{\rm \theta}(r,\theta)=-B_{\rm
0}\frac{3\phi(\theta)}{8\pi\sin\theta}\frac{1}{(r/r_{\rm
0})^{5/4}},
\end{equation}
\begin{equation}
\ B_{\varphi}(r,\theta)=B_{\rm 0}\frac{b(\theta)}{(r/r_{\rm
0})^{5/4}},
\end{equation}
where $\rho_{\rm 0}$, $p_{\rm 0}$, $B_{\rm 0}$ and $r_{\rm 0}$
provide convenient units with which the equations can be written
in non-dimensional forms and $\Omega_{\rm K}$ denotes the
Keplerian angular velocity, viz. $\Omega_{K}(r)=\sqrt{GM/r^3}$.
Also, we obtain the following equations for the components of
current density:
\begin{equation}
\ J_{\rm r}(r,\theta)=\frac{B_{\rm 0}}{r_{\rm
0}}\frac{1}{\sin\theta}\frac{d}{d\theta}(b(\theta)\sin\theta)\frac{1}{(r/r_{\rm
0 })^{9/4}},
\end{equation}
\begin{equation}
\ J_{\rm\theta}(r,\theta)=\frac{B_{\rm 0}}{4r_{\rm
0}}\frac{b(\theta)}{(r/r_{\rm 0})^{9/4}},
\end{equation}
\begin{equation}
\ J_{\rm\varphi}(r,\theta)=\frac{B_{\rm 0}}{2\pi r_{\rm
0}}[\frac{3\phi(\theta)}{16\sin\theta}-
\frac{d}{d\theta}(\frac{1}{\sin\theta}\frac{d\phi(\theta)}{d\theta})]\frac{1}{(r/r_{\rm
0 })^{9/4}},
\end{equation}
and considering equation (\ref{eq:diffu}) magnetic diffusivity
$\eta$ becomes
\begin{equation}
\eta=\eta_{\rm 0}\frac{p_{\rm 0}}{\rho_{\rm 0}\sqrt{GM/r_{\rm
0}^3}}\frac{P(\theta)}{R(\theta)}(\frac{r}{r_0})^{1/2}.
\end{equation}

Substituting the above solution in the momentum, induction and
energy equations, we obtain a set of coupled ordinary
differential equations. Note that the above solutions
automatically satisfy continuity equation (\ref{eq:continuty}).
Equations of motion (\ref{eq:motionr}), (\ref{eq:motiontet}) and
(\ref{eq:motionphi}) become
\begin{displaymath}
\ R(-\frac{V^{2}}{2}-\Omega^{2})=\frac{5}{2}A_{\rm 1} P - R +
\frac{A_{\rm 2}}{16\pi}\{b^{2}+
\end{displaymath}
\begin{equation}
\frac{3\phi}{4\pi^{2}\sin\theta}[\frac{3\phi}{16\sin\theta}-\frac{d}{d\theta}(\frac{1}{\sin\theta}\frac{d\phi}{d\theta})]\},
\label{eq:m1}
\end{equation}
\begin{displaymath}
-R\Omega^{2}\cot\theta=-A_{1}\frac{dP}{d\theta}+\frac{A_{2}}{4\pi}\{\frac{1}{4\pi^{2}\sin\theta}\frac{d\phi}{d\theta}\times
\end{displaymath}
\begin{equation}
[\frac{3\phi}{16\sin\theta}-\frac{d}{d\theta}(\frac{1}{\sin\theta}\frac{d\phi}{d\theta})]-\frac{b}{\sin\theta}\frac
{d}{d\theta}(b\sin\theta)\}, \label{eq:m2}
\end{equation}
\begin{equation}
\
RV\Omega=-\frac{A_{2}}{4\pi^{2}\sin\theta}[b\frac{d\phi}{d\theta}+\frac{3\phi}{4\sin\theta}\frac{d}{d\theta}(b\sin\theta)].
\label{eq:m3}
\end{equation}
Equations (\ref{eq:IIII}) gives
\begin{displaymath}
\ 2\pi V
b-\frac{d}{d\theta}(\Omega\phi)-\frac{\Omega}{\sin\theta}\frac{d\phi}{d\theta}+
\end{displaymath}
\begin{equation}
\frac{\eta_{0}}{2}\frac{Pb}{R}+\frac{8\pi\eta_{\rm
0}}{3}\frac{d}{d\theta}[\frac{P}{R\sin\theta}\frac{d}{d\theta}(b\sin\theta)]=0,
\label{eq:m5}
\end{equation}
and equation (\ref{eq:IV}) is cast into this form
\begin{equation}
\frac{d}{d\theta}(\frac{1}{\sin\theta}\frac{d\phi}{d\theta})-\frac{3\phi}{16\sin\theta}-\frac{3}{4\eta_{\rm
0}A_{\rm 1}} \frac{RV\phi}{P\sin\theta}=0, \label{eq:m4}
\end{equation}
and finally the energy equation (\ref{eq:energyr}) becomes
\begin{displaymath}
\ RV\frac{3\gamma-5}{2(\gamma-1)}=\frac{f\eta_{\rm
0}A_{2}}{4\pi}\{\frac{1}{4\pi^{2}}[\frac{3\phi}{16\sin\theta}-\frac{d}{d\theta}(\frac{1}{\sin\theta}\frac{d\phi}{d\theta})]^{2}+
\end{displaymath}
\begin{equation}
\frac{b^{2}}{16}+[\frac{1}{\sin\theta}\frac{d}{d\theta}(b\sin\theta)]^{2}\},
\label{eq:m6}
\end{equation}
where
\begin{equation}
\ A_{1}=\frac{p_{\rm 0}}{\rho_{\rm 0}(GM/r_{\rm
0})}=(\frac{c_{\rm 0s}}{v_{\rm 0K}})^2,
\end{equation}
\begin{equation}
A_{2}=\frac{B_{\rm 0}^{2}}{\rho_{\rm 0}(GM/r_{\rm 0})}=4\pi
(\frac{v_{\rm 0A}}{v_{\rm 0K }})^2,
\end{equation}
where $v_{\rm 0K}=(GM/r_{\rm 0})^{1/2}$ is the Keplerian velocity,
and $v_{\rm 0A}=B_{\rm 0}/\sqrt{4\pi \rho_{\rm 0}}$ is
Alfv$\acute{e}$n velocity.

We can define the ratio of the thermal to the magnetic pressures
$\beta$ as
\begin{equation}
\beta(\theta)=\beta_{\rm 0}\frac{4\pi^2 P \sin^2\theta
}{9\phi^2/16+(d\phi/d\theta)^2+4\pi^2 b^2\sin^2\theta},
\end{equation}
where $\displaystyle \beta_{\rm 0}=\frac{p_{\rm 0}}{B_{\rm
0}^{2}/8\pi}=8\pi \frac{A_{\rm 1}}{A_{\rm 2}}$.
Equations (\ref{eq:m1})-(\ref{eq:m6}) constitute a system of
ordinary non-linear differential equations for the six similarity
variables $\phi(\theta)$, $b(\theta)$, $R(\theta)$, $P(\theta)$,
$V(\theta)$ and $\Omega(\theta)$. Indeed, the behaviour of the
solution depends on boundary conditions which are supposed based
on some physical assumptions such as symmetry with respect to the
equatorial plane. Ogilvie (1997) studied equilibrium of
magnetized non-accreting disks by assuming self-similarity in
spherical radial coordinate. If we neglect resistivity, our
solution reduces to the Ogilvie's solution, i.e. the model
describes a non-accreting disk containing poloidal and toroidal
components of magnetic field. However, in the case of resistive
accretion flow, the scalings of quantities with spherical radius
are still similar to those presented by Ogilvie (1997) and some
authors (e.g., Blandford \& Payne 1982), i.e. $\rho \propto
r^{-3/2}$, $p\propto r^{-5/2}$ and ${\bf B} \propto r^{-5/4}$.

The main focus of this study is nonrotating accretion flow. But an
interesting feature of the solutions is worth emphasizing. Using
equation (\ref{eq:m4}), we can rewrite equation (\ref{eq:m1}) as
\begin{equation}
\\R(1-\frac{V^2}{2}-\Omega^2)=\frac{5}{2}A_{\rm 1}P+\frac{A_{\rm 2
}}{16\pi}(b^2-\frac{9RV \phi^2}{16\pi^2 \eta_{\rm 0} P
\sin^2\theta}). \label{eq:m7}
\end{equation}
Since all the similarity variables are positive except for $V$,
we see that the right hand side of the equation (\ref{eq:m7}) is
positive for all $\theta$ except for $\theta=0$ which is not
evident from the equation. For nonrotating flow within purely
poloidal magnetic field (i.e., $\Omega=0$ and $b=0$) which is our
interest in this paper, if we substitute from equations
(\ref{eq:m4}) and (\ref{eq:m6}) into the last term of right hand
side of the above equation, it is positively well-defined for all
$\theta$, even in the limit of vanishing $\theta$. However, if we
consider both the toroidal and the poloidal  components of
magnetic field, we can still say that when $\theta$ tends to
zero, the last term of right hand side of equation (\ref{eq:m7})
even if tends to a very large value, it would be negative
(because $V$ is negative, the rest of the variables are positive)
and so, the right hand side of the equation is positive (there is
minus sign before the last term of right hand side). Thus, we
should have
\begin{equation}
\frac{V^2}{2}+\Omega^2 \leq 1.
\end{equation}
This inequality implies rotation velocities at below Keplerain
rate for the solutions of the model, irrespective of the amount of
cooling or value of resistivity $\eta_{\rm 0}$. Also, we can say
as the value of angular velocity increases, the accretion velocity
decreases. In other words, in this model, the high accretion rate
corresponds to low rotation rate. In this study, we shall
restrict our attention to non-rotating accretion flow, i.e.
$\Omega=0$.

\subsection{Nonrotating accretion  Solutions}

What we have is a set of complicated differential equations which
must be solved under appropriate boundary conditions. Although a
full numerical solutions to the equations
(\ref{eq:m1})-(\ref{eq:m6}) would now be possible, it is more
instructive to proceed by analyzing  the model in some
restrictive conditions such as nonrotating case. To derive this
solution we set $\Omega=0$ and $b=0$, which the later condition
means that we consider purely poloidal magnetic fields. However,
to further simplify the problem, we can not impose spherical
symmetry, i.e. $d/d\theta\equiv 0$. Because if we impose
spherical symmetry, we can not identify the actual magnetic field
lines, nor can we account for nonradial forces. But we know that
in resistive accretion flow the topology of magnetic field has a
vital role.

The same as NY we are defining  parameters $\displaystyle
\epsilon=\frac{5/3-\gamma}{\gamma-1}$ and $\displaystyle
\epsilon_{\rm f}=\frac{\epsilon}{f}$. As we will see the
parameter $\epsilon_{\rm f}$ plays an important role in the
behaviour of the solutions. Obviously, when we set $\Omega=0$ and
$b=0$, equations (\ref{eq:m3}) and (\ref{eq:m5}) then immediately
drop out and the rest of equations, i.e equations (\ref{eq:m1}),
(\ref{eq:m2}) and (\ref{eq:m6}) can be written as

\begin{displaymath}
\ R(1-\frac{V^2}{2})=\frac{5}{2}A_{\rm 1}P-\frac{3A_{\rm
2}}{64\pi^3}\times
\end{displaymath}
\begin{equation}
\frac{\phi}{\sin\theta}[\frac{d}{d\theta}(\frac{1}{\sin\theta}\frac{d\phi}{d\theta})-
\frac{3\phi}{16\sin\theta}],\label{eq:denv1}
\end{equation}
\begin{equation}
\ A_{\rm 1}\frac{dP}{d\theta}+\frac{A_{\rm
2}}{16\pi^{3}\sin\theta}[\frac{d}{d\theta}(\frac{1}{\sin\theta}\frac{d\phi}{d\theta})-
\frac{3\phi}{16\sin\theta}]\frac{d\phi}{d\theta}=0,\label{eq:denv2}
\end{equation}
\begin{equation}
\ [\frac{d}{d\theta}(\frac{1}{\sin\theta}\frac{d\phi}{d\theta})-
\frac{3\phi}{16\sin\theta}]^{2}+\frac{24\pi^3\epsilon_{\rm
f}}{\eta_{\rm 0}A_{\rm 2}}RV=0, \label{eq:densv}
\end{equation}
and equation (\ref{eq:m4}) remains unchanged. Now, the set of
equations reduces to four equations (\ref{eq:m4}),
(\ref{eq:denv1}), (\ref{eq:denv2}) and (\ref{eq:densv})   which we
want to solve.

Clearly the expressions inside the brackets in equations
(\ref{eq:denv1}), (\ref{eq:denv2}) and (\ref{eq:densv}) are
similar and by substituting from equation (\ref{eq:m4}), these
three equations become
\begin{equation}
\ R(1-\frac{V^2}{2})=\frac{5}{2}A_{1}P-\frac{9A_{2}}{256\pi^3
\eta_{0}A_{1} }\frac{RV\phi^2}{P\sin^2\theta},\label{eq:mi1}
\end{equation}
\begin{equation}
\ A_{1}\frac{dP}{d\theta}+\frac{3A_{2}}{64\pi^3\eta_{0}A_{1}}
\frac{RV\phi}{P\sin^{2}\theta}\frac{d\phi}{d\theta}=0,\label{eq:mi2}
\end{equation}
\begin{equation}
\frac{RV\phi^2}{P^2\sin^2\theta}+\frac{128\pi^3\eta_{0}\epsilon_{\rm
f } A_{1}^2}{3A_{2}}=0.\label{eq:mi3}
\end{equation}
From equation (\ref{eq:mi3}), we obtain
\begin{equation}
\ RV=-\frac{128\pi^3\eta_{0}\epsilon_{\rm f } A_{1}^2}{3A_{2}}
\frac{P^2 \sin^2\theta}{\phi^2},\label{eq:rv}
\end{equation}
and by substituting $RV$ from the above equation into equations
(\ref{eq:mi1}) and (\ref{eq:mi2}), we have
\begin{equation}
\ R(1-\frac{V^2}{2})=(\frac{5+3\epsilon_{\rm f}}{2})A_1
P,\label{eq:mi4}
\end{equation}
\begin{equation}
\frac{1}{P}\frac{dP}{d\theta}+\frac{2\epsilon_{\rm
f}}{\phi}\frac{d\phi}{d\theta}=0.\label{eq:mi5}
\end{equation}
Fortunately, equation (\ref{eq:mi5}) is integrable and gives
\begin{equation}
\ P=\phi^{n}, \label{eq:pressure}
\end{equation}
where $n=2\epsilon_{\rm f}$. Now, considering the above result,
we can rewrite equations (\ref{eq:rv}) and (\ref{eq:mi4}) as
\begin{equation}
\ RV = -\frac{128\pi^3\eta_{0}\epsilon_{\rm f } A_{1}^2}{3A_{2}}
\phi^{2n-2} \sin^2\theta,\label{eq:J1}
\end{equation}
\begin{equation}
\ R(1-\frac{V^{2}}{2})=(\frac{5+3\epsilon_{\rm f}}{2})A_1
\phi^n.\label{eq:J2}
\end{equation}
From these two equations, one can easily calculate $V$ and $R$ as
functions of $\phi$ as follows
\begin{equation}
\ V(\theta)=\frac{1-\sqrt{1+2 a^{2}
\phi^{2(n-2)}\sin^{4}\theta}}{a\phi^{n-2}\sin^{2}\theta},
\label{eq:v1}
\end{equation}
where
\begin{equation}
\ a=\frac{32\pi^{2}\beta_{\rm 0}\eta_{\rm 0}\epsilon_{\rm
f}}{3(5+3\epsilon_{\rm f})},
\end{equation}
and by substituting $V$ as function of $\phi$ into equations
(\ref{eq:J1}) or (\ref{eq:J2}), we can find $R$ as function of
$\phi$. Thus, if we can find $\phi$, all physical variables are
known. Equation (\ref{eq:m4}) is the last equation which gives us
$\phi$ as a function of $\theta$. Considering equations
(\ref{eq:pressure}) and (\ref{eq:J1}), we can rewrite equation
(\ref{eq:m4}) as
\begin{equation}
\frac{d}{d\theta}(\frac{1}{\sin\theta}\frac{d\phi}{d\theta})-\frac{3\phi}{16\sin\theta}+4\pi^{2}\beta_{\rm
0}\epsilon_{\rm f}\phi^{n-1}\sin\theta=0
\end{equation}
or,
\begin{equation}
\frac{d^{2}\phi}{d\xi^{2}}-\frac{3\phi}{16(1-\xi^{2})}+4\pi^{2}\beta_{\rm
0}\epsilon_{\rm f}\phi^{n-1}=0, \label{eq:main}
\end{equation}
where $\xi=\cos \theta$. Now, we have a second order ordinary
differential equation for $\phi(\theta)$ which can be solved
using a numerical method (in Appendix we can find an analytical
solution for $n=1$). One boundary condition can be specified on
the equatorial plane, by assuming that field lines thread the
equator vertically:
\begin{equation}
\ (\frac{d\phi}{d\theta})_{\theta=\pi/2}=0.
\end{equation}
The second boundary condition can be specified by demanding the
magnetic flux enclosed by the polar axis to vanish:
\begin{equation}
\phi(0)=0.
\end{equation}

 \begin{figure}
   \centering
   \includegraphics[width=9cm]{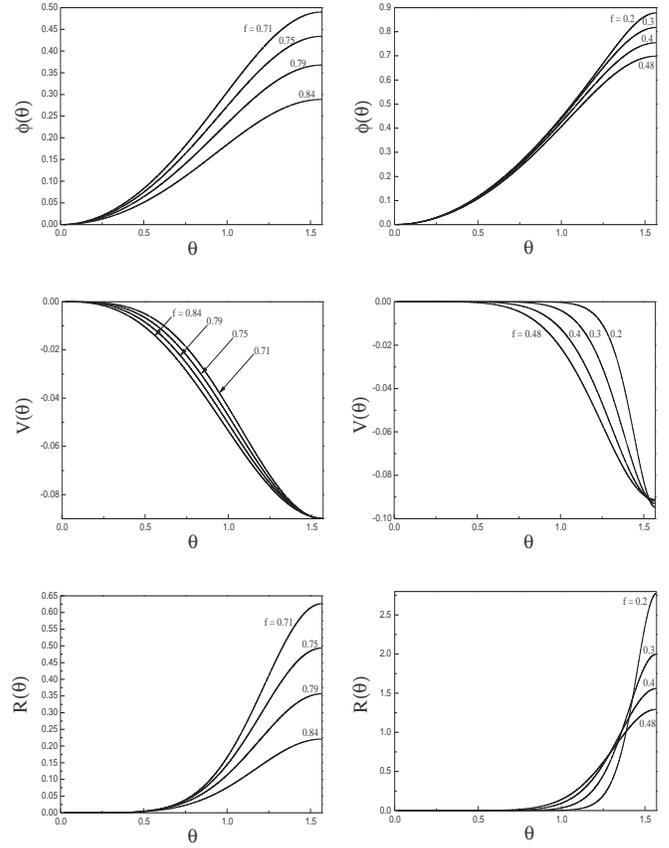}
      \caption{Self-similar solutions corresponding to
      $\gamma=4/3$, $\eta_{0}=0.1$ and $\beta_0=0.1$. Top:
      magnetic flux function $\phi(\theta)$ as a function of polar
      angel $\theta$. Middle: Radial velocity $V(\theta)$. Bottom:
      density $R(\theta)$.
              }
         \label{fig1}
   \end{figure}
Thus, the complicated set of differential equations is reduced to
a boundary value problem of a second order differential equation.
We have obtained numerical solutions of the above equation for
different values of $f$ and $\eta_{\rm 0}$. For our illustrative
parameters, we assume $r_{\rm 0}= 1 $pc , $B_{\rm 0 }=10^{-5}$G,
$M=10^6$ M$_{\odot}$, $\beta_{\rm 0}=0.1$ and $\dot{M}_{\rm
0}=2\pi \rho_{\rm 0} r_{\rm 0}^2 \sqrt{GM/r_{\rm 0}}=1.2\times
10^{-4}$ M$_{\odot}$/year which gives $\rho_{\rm 0}=2\times
10^{-20}$ g/cm$^3$. Figure \ref{fig1} shows some particular
solutions corresponding to $\eta_{\rm 0}=0.1$, $\gamma=4/3$ and
different values of the advected energy $f$. For a fixed value of
$\gamma$, the solutions imply increasing cooling. This general
behavior is the same as NY for viscous ADAF.

The behaviour of the solutions, as the six panels in Figure
\ref{fig1} show, is very sensitive to the value of $f$ or
$\epsilon_{\rm f}$. The top  panels show variations of magnetic
flux function $\phi(\theta)$ as a function of $\theta$. While
this similarity magnetic flux is zero on the polar axis (this is
a boundary condition), it has a maximum value on the equator. The
maximum value of $\phi(\theta)$ on the equator increases with
decreasing $f$. This growing flux contrast between the equatorial
region and the polar region implies that at a fixed radius on the
equator, the magnetic flux increases with decreasing $\gamma$
(or, equivalently, increasing cooling).

The middle panels show similarity function of radial velocity
$V(\theta)$ (see equation (\ref{eq:v1})). We see that the maximum
accretion velocity is at equatorial region and on the polar axis
there is no mass inflow. As we expected the accretion velocity is
sub-Keplerian. Although for our representative set of parameters
the variation of $V(\theta)$ as a function of $\theta$ is
decreasing from equatorial region to polar region, we can
recognize  two distinct regions in the general profile of
$V(\theta)$ depending on the value of the velocity. The bulk of
accretion occurs from the equatorial plane at $\theta=\pi/2$ to a
surface at $\theta=\theta_{\rm f}$, inside of which the velocity
$V(\theta)$ is zero. Note that all of our solutions have
well-behaved nonsigular $V(\theta)$ at all $\theta$ and the
radial velocity does not diverge at $\theta=\theta_{\rm f}$.
Also, for the flows which extend from the equatorial plane to a
surface at $\theta=\theta_{\rm s}$ where $\theta_{\rm
f}<\theta_{\rm s}<\pi/2$,  the accretion velocity is decreasing.
With decreasing $f$, the region with maximum accretion velocity
becomes thin, i.e. $\theta_{\rm s}$ decreases.

In the bottom  panels of Figure \ref{fig1}, we plot the density
similarity function $R(\theta)$. The density contrast between
equatorial and polar regions increases with decreasing $f$. The
same as the magnetic flux function $\phi(\theta)$, as $f$
decreases, the density grows and becomes concentrated toward the
equatorial plane. As already mentioned, the small $f$ solutions
correspond to efficient cooling for a fixed $\gamma$. Thus, the
$R(\theta)$ profile clearly shows what we expect with increasing
cooling, or equivalently $\epsilon_{\rm f}$. More interestingly,
although the solutions for the large $\epsilon_{\rm f}$ implies
the bulk of the accretion occurs in equatorial region (i.e., like
a thin disk configuration), in all cases there is a low density
with higher temperature above the disk as the bottom panels of
Figure \ref{fig1} shows. While bulk  of the accreting gas along
the equatorial plane has approximately constant temperature, this
tenuous region also has a constant higher temperature. As cooling
increases, the temperatures of both regions decrease. Note that
at tenuous region, there is no accretion.
\begin{figure}
   \centering
   \includegraphics[width=8cm]{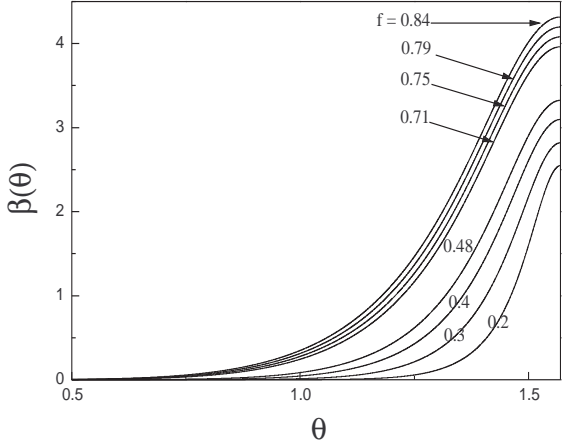}
      \caption{Ratio of the thermal to the magnetic pressures for
      the solutions shown in Figure \ref{fig1}.
              }
         \label{fig2}
   \end{figure}

From these Figures we see for fixed values of $\gamma$ and
$\eta_{\rm 0}$, as the advected energy decreases and more energy
radiates out of the system, magnetic flux function $\phi(\theta)$
increases. It means magnetic field lines can penetrate through
the disk easier, when we have cooling. Thus,  while thermal
pressure decreases due to the cooling, the magnetic pressure
increases and we have lower values for the ratio of these two
which seems to be important in analyzing the dynamics of the flow.
\begin{figure}
   \centering
   \includegraphics[width=9cm]{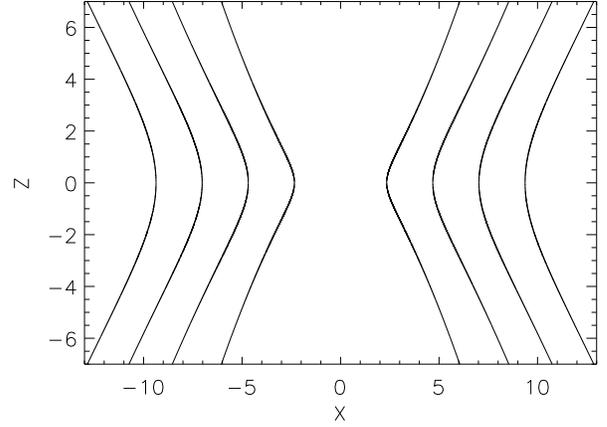}
      \caption{Representation of magnetic field lines in the
      meridional plane for a typical self-similar solution with
      $\eta_0=0.1$, $\gamma=4/3$ and $f=0.84$.
              }
         \label{fig3}
   \end{figure}
All solutions correspond to poloidal magnetic field
configuration, in which the field lines bend only once when
passing through the equatorial plane.

Figure \ref{fig2} shows the ratio of the thermal to magnetic
pressures. We see that this ratio decreases by decreasing the
amount of the advected energy $f$. Figure \ref{fig3} shows
magnetic field lines for a typical self-similar solution with
$\eta_{\rm 0}=0.1$ and $ f=0.84$. For other input parameters, we
can obtain the same configuration. Ogilvie (1997) studied
nonaccreting magnetized disk by considering a polytropic equation
of state. However, the magnetic field configuration in his study
is similar to what has been obtained in this study for resistive
accretion flow. If we neglect turbulent diffusivity and the
energy equation, our solutions reduce to Ogilvie's solutions.
But, in our scenario, two main factors in dynamics of the flow
are the diffusion mechanism of magnetic field lines and the
advected energy. For example, the net rate of accretion depends
on the enclosed flux function $\phi(\theta)$. We can write
\begin{equation}
\dot{M}=-2\dot{M}_{\rm 0}\int_{0}^{\frac{\pi}{2}}\sin\theta
R(\theta) V(\theta) d\theta.
\end{equation}
Considering equations (\ref{eq:densv}) and (\ref{eq:pressure}), we
can rewrite the above equation as
\begin{equation}
\dot{M}=\frac{16}{3}\pi^2 \beta_{\rm 0} n \eta_{0}
A_{1}\dot{M}_{\rm 0}\int_{\rm
0}^{\frac{\pi}{2}}\phi^{2n-2}\sin^3\theta d\theta.
\end{equation}
Although this kind of magnetized accretion flow is different from
Bondi (1952) accretion in various aspects, we can define Bondi
accretion rate for the illustrative parameters as $\dot{M}_{\rm
Bondi}\simeq 4\pi G^2 M^2 \rho_{\rm 0}/c_{\rm s}^3$. Thus,
$\dot{M}_{\rm Bondi}=2(v_{\rm 0K}/c_{\rm 0s})^3 \dot{M}_{\rm 0}$.
Assuming $A_{\rm 1}=1$, we  can write the net accretion rate as
follows
\begin{equation}
\dot{M}=\frac{8}{3}\pi^2 \beta_{\rm 0} n \eta_{0}\dot{M}_{\rm
Bondi} \int_{\rm 0}^{\frac{\pi}{2}}\phi^{2n-2}\sin^3\theta
d\theta.
\end{equation}
\begin{figure}
   \centering
   \includegraphics[width=8cm]{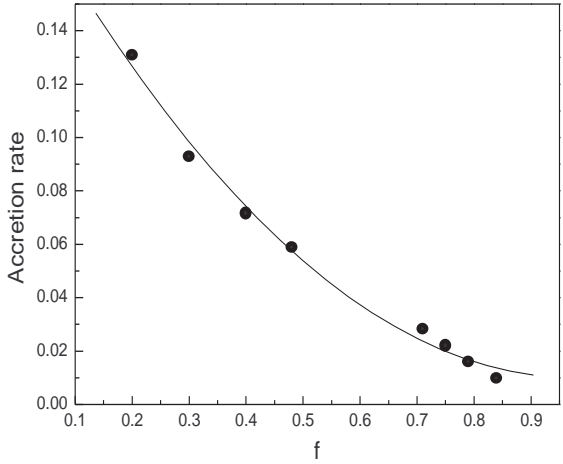}
      \caption{The net accretion rate $\dot{M}/\dot{M}_{\rm
      Bondi}$ as a function of the advected energy $f$ for
      self-similar solutions shown in Figure \ref{fig1}.
              }
         \label{fig4}
   \end{figure}
For the flows which we have at hand,  the mass accretion rate is
at below Bondi accretion rate for wide range of input parameters
(e.g., $\epsilon_{\rm f}$ and $\eta_{\rm 0}$). Figure \ref{fig4}
shows the net mass accretion rate $\dot{M}/\dot{M}_{\rm Bondi}$
as a function of the advected energy $f$ corresponding to $f=0.84,
0.79, 0.75, 0.71, 0.48, 0.4, 0.3, 0.2$ and $\gamma=4/3$ and
$\eta_{0}=0.1$. As the above equation shows the mass accretion
rate changes in proportion to $\eta_0$. For these parameters, we
have fitted an approximate function as $\dot{M}/\dot{M}_{\rm
Bondi}\approx 0.19-0.38f+0.19f^2$. The radial structure of the
resistive radiatively inefficient accretion flow is dramatically
modified from Bondi profile, and the mass accretion rate is
decreased significantly by increasing the advected energy $f$. We
see for a fixed value of $\eta_{\rm 0}$, the enclosed flux
function $\phi(\theta)$ increases as $\epsilon_{\rm f}$ increases
(i.e, cooling increases) and the net mass accretion rate
increases, as well. Thus, in our model, the accretion rate is
always suppresses comparing to the Bondi rate, although the
precise amount depends on the cooling and the resistivity. There
are many numerical simulations which show the infall of
non-rotating or slowly-rotating magnetized gas with reduced
$\dot{M}$ (e.g., Igumenshchev et al. 2000; Hawley, Balbus \&
Stone 2001). However, our simple model for magnetized accretion
flow shows this behavior clearly.

\section{Discussion and Conclusions}

In this paper, we have obtained axisymmetric magnetized
self-similar advection-dominated flow solutions by considering an
ordered magnetic field and a resistive dissipation of the energy
of the flow. The radial dependence of the solutions has the same
behaviour as ADAF solutions (NY), but the angular parts are
obtained by  solving a set of ordinary differential equations. In
the model, the viscosity of the flow is completely neglected in
order to make the contrast between the standard viscous  and
resistive ADAF solutions clear. We showed that the radial and the
rotational velocities are well below the Keplerian velocity, and
this is independent of the resistivity and the amount of cooling.
However, the main aim of this study was to study the nonrotating
quasi-spherical magnetized flow directly just by solving the
relevant MHD equations. The same problem has been addressed by
many authors, under the assumption that the field will achieve
equipartition of magnetic and kinetic energies (see, e.g.,
Shvartsman 1971; NY;  Bisnovatyi-Kogan and Lovelace 2000).
Bisnovatyi-Kogan and Lovelace (2000) have  discussed about
equipartition between magnetic and kinetic energies using
time-averaged magnetic field solution. They argued for such an
equipartition half of the dissipated energy of the accretion flow
results from the destruction of the magnetic field. In this paper
we showed that the equipartition depends on the input parameters,
more importantly the amount of advected energy ($\epsilon_{\rm
f}$) and the resistivity $\eta_{\rm 0}$. Also, the unmagnetized
problem is described by pure hydrodynamics and it was solved by
Bondi (1952) and has been widely applied. Note that our solutions
differ in various aspects from Bondi (1952) spherical accretion.
While the magnetic field diffusivity is the main factor in
accretion in our model, the energy equation has been considered
as well. In Bondi accretion, there is no mechanism for energy
transport. Note that in pure Bondi spherical flow, a self-similar
accretion is allowed only for a single value of $\gamma$.
However,  magnetized rediatively inefficient accretion flow is
possible for a range of values of $\gamma$. The other interesting
feature is that the dynamics of such flow depends not only to
magnetic diffusivity $\eta_{\rm 0}$, but also the fraction of
advected energy $f$ is important.

For a nonrotating resistive accretion flow, the system of
equations  simplified to a second order differential equation.
For given $\epsilon_{\rm f}$ and $\eta_{\rm 0}$, this equation
determines the magnetic field and the other physical variables can
be found using a set of obtained algebraic equations. Although,
we have neglected ingredients such as angular momentum and
viscosity, the general properties of our solutions are similar to
ADAF solutions. More importantly, the geometrical shape of the
flow is determined by the amount of the advected energy (i.e.,
$\epsilon_{\rm f}$) for a given $\eta_{\rm 0 }$. While solutions
with efficient cooling resemble to thin disks, the radiatively
inefficient accretion describe nearly quasi-spherical flows.

The bulk of accretion with nearly constant velocity occurs in a
region which extends from the equatorial plane to a surface at
$\theta_{\rm s}$, inside of which the accretion velocity
gradually decreases to zero. Thus, the accretion velocity is
constant in the region $\pi/2 <\theta<\theta_{\rm s}$. As
$\epsilon_{\rm f}$ decreases, the size of this region increases
and the accretion tends to a quasi-spherical flow. The
temperature of this region is constant as well and is surrounded
by a region with low-density and higher temperature. This hot gas
corona is similar to what has been discussed in viscous ADAF
solutions (NY). As the cooling increases and the flow tends to
disk like configuration, the size of this corona increases.

The ratio of the thermal to the magnetic pressure $\beta(\theta)$
depends on the input parameters $\epsilon_{\rm f}$ and $\eta_{\rm
0 }$. In the limit of large $\epsilon_{\rm f}$, the value of
$\beta(\theta)$ is lower than one for all $\theta$. It means that
the thermal pressure is not in equipartition with the magnetic
pressure and more precisely, the magnetic pressure is stronger
than thermal pressure in this regime. On the other hand, for
inefficient cooling flows, we found that the value of the ratio
exceeds than one for a range of $\theta$. Still there is no
 equipartition between the thermal and the magnetic pressures
 except for a surface at a specific angle which can be
 determined.

As for the rate of heating $Q_{\rm diss}$, the resistive
dissipation at small scales as a result of a turbulence cascade
has been assumed as the dominant mechanism. However, there are
other sources such as energy release through magnetic
reconnection and more importantly viscous dissipation.
Phenomenological considerations may lead us to define $Q_{\rm
diss}$ as a fraction of the gravitational potential energy of the
accreting gas. This simple definition allows one to consider all
the dissipation processes without having detailed models for
them. Preliminary results for this form of $Q_{\rm diss}$ shows
that the scalings of quantities with radius still remains similar
to our model, but the other properties of the solutions should be
studied.

Since we completely neglected the viscous dissipation, the
turbulent magnetic Prandtel number of our model is zero. This
unsatisfactory aspect can be removed by considering both the
viscosity and the resistivity simultaneously. Fortunately, the
radial scaling of the physical variables in the resistive case,
as we showed, is similar to the  viscous case. By dimensional
analysis, one can simply show that self-similarity is possible as
long as $\eta$ (and viscosity $\nu$) scales with radius as
$r^{1/2}$. Thus, it is still possible to find self-similar
solutions in the general case. Given the importance of the value
of the turbulent magnetic Prandtel number, we can construct a
model for viscous-resistive accretion flow, in which this number
is a free parameter.


\appendix\section[]{}
Equation (\ref{eq:main}) can be solved analytically for $n=1$.
Assuming
\begin{equation}
\phi(\xi)=\frac{16}{35}4\pi^{2}\beta_{\rm 0}\epsilon_{\rm
f}(1-\xi^2)+y(\xi),
\end{equation}
equation (\ref{eq:main}) reduces to
\begin{equation}
\frac{d^2y}{d\xi^2}-\frac{3}{16}\frac{y}{1-\xi^2}=0.
\end{equation}
By changing the variables as
\begin{equation}
\ z=\frac{1+\xi}{2},  y=zY
\end{equation}
this equation becomes
\begin{equation}
\ z(1-z)\frac{d^2Y}{dz^2}+(2-2z)\frac{dY}{dz}-\frac{3}{16}Y=0.
\end{equation}
The general solution of this hypergeometric equation is
\begin{displaymath}
\ Y(z)=C_{\rm 1}F(\frac{3}{4},\frac{1}{4}|2|z)+C_{\rm
2}F(\frac{3}{4},\frac{1}{4}|2|z)\times
\end{displaymath}
\begin{equation}
\int\frac{dz}{[zF(\frac{3}{4},\frac{1}{4}|2|z)]^2},
\end{equation}
where $C_{\rm 1}$ and $C_{\rm 2}$ are arbitrary constants and $F$
is hypergeometric function.

\label{lastpage}

\end{document}